\documentclass[article, aps,prd,a4paper,superscriptaddress,tightenlines,noeprint,11pt,floatfix]{revtex4-2}

\usepackage{amsmath,amssymb,amsfonts}
\usepackage{braket}
\usepackage{bbm}
\usepackage{xcolor,ulem}
\usepackage{graphicx}
\usepackage{hyperref}
\usepackage{cleveref}

\newcommand{\ignore}[1]{}
\newcommand{\be}{\begin{equation}}
\newcommand{\ee}{\end{equation}}
\def\norm#1{\Vert #1\Vert}
\newcommand{\ba}{\begin{eqnarray}}
\newcommand{\ea}{\end{eqnarray}}
\newcommand{\hil}{\mathcal{H}}
\newcommand{\mg}{\mathcal{G}}
\newcommand{\hi}{\mathcal{H}}
\DeclareMathOperator{\tr}{tr}
\newcommand\stab{{\operatorname{PSTAB}}}

\newcommand{\ot}{\otimes}
\newcommand{\sswap}{{\rm T}}
\newcommand{\ketbra}[1]
{\ket{#1}\hspace{-0.1 cm} \bra{#1}}

\newcommand{\absval}[1]{\left| {#1} \right|}
\newcommand{\Id}{\mathbbm{1}}
\newcommand{\cg}{|\mathcal{G}|}
\newcommand{\id}{\mathbbm{1}}

\newcommand{\sym}{\text{sym}}

\newcommand{\exv}{\mathbb{E}}
\newcommand{\lin}{\operatorname{lin}}

\newcommand{\mlin}{M_{\lin}}
\newcommand{\pauli}{\mathbb{P}}
\newcommand{\ttt}{\sswap}

\begin{document}
\title{Magic of discrete lattice gauge theories
}

\author{Gianluca Esposito}

\affiliation{Scuola Superiore Meridionale, , Largo S. Marcellino 10, 80138 Napoli, Italy \\
 INFN, Sezione di Napoli, Italy \\
\email{g.esposito@ssmeridionale.it} }

\author{Simone Cepollaro}

\affiliation{Scuola Superiore Meridionale, , Largo S. Marcellino 10, 80138 Napoli, Italy \\
 INFN, Sezione di Napoli, Italy \\
\email{simone.cepollaro-ssm@unina.it} }

\author{Luigi Cappiello}

\affiliation{INFN, Sezione di Napoli, Italy \\
 Dipartimento di Fisica `Ettore Pancini', Universit\`a degli Studi di Napoli Federico II,
Via Cintia 80126,  Napoli, Italy \\
\email{luigi.cappiello@na.infn.it} }

\author{Alioscia Hamma}

\affiliation{Dipartimento di Fisica `Ettore Pancini', Universit\`a degli Studi di Napoli Federico II, Via Cintia 80126,  Napoli, Italy\\
NFN, Sezione di Napoli, Italy \\
Scuola Superiore Meridionale, Largo S. Marcellino 10, 80138 Napoli, Italy  \\ 
\email{alioscia.hamma@unina.it} }

\begin{abstract}
Simulation of quantum field theories and fundamental interactions are one of the most challenging tasks in modern particle physics. Classical computers generally fail to reproduce accurate results when it comes to strongly coupled theories such as QCD. Recent developments in quantum technologies open up the possibility of simulating such physical regimes by using quantum computers. In this paper, we study the quantum resource related to the simulability of a quantum theory, i.e. non-stabilizerness for Lattice Gauge Theory (LGT) with discrete symmetry gauge groups. We show that enforcing gauge constraints for $\mathbb{Z}_l$ LGTs has no cost in terms of this resource and discuss the relation between non-abelianity of the gauge group with the average non-stabilizerness of the gauge invariant Hilbert space.     
\end{abstract}

\maketitle

\section{Introduction}
Gauge theories describe fundamental interactions through the principle of local gauge invariance. This principle requires that the laws of physics remain unchanged under certain local transformations of the fields, leading to the existence of gauge bosons that mediate the forces.
The development of gauge theories and the Standard Model represents one of the most significant achievements in twentieth-century physics, culminating in a unified description of fundamental particles and forces. The concept of gauge invariance emerged from Weyl's attempt to unify electromagnetism and general relativity \cite{Weyl1929}. Although unsuccessful, his work laid the foundation for future developments. In particular, the quantum revolution led to the development of quantum mechanics and early quantum field theory.  However,  in quantum field theory, matter and gauge fields, the mediators of the interactions, are described by infinite numbers of degrees of freedom and this leads almost immediately to the appearance of divergencies in the calculations of perturbative corrections to physical quantities. The solution to the problem of infinities was obtained in the 50s by Tomonaga \cite{Tomonaga1946}, Schwinger \cite{Schwinger1948}, and Feynman \cite{Feynman1949} who independently developed renormalized QED, establishing it as the first successful quantum gauge theory, based on the Abelian group $U(1)$. 

The discovery of new particles led to the application of non-Abelian gauge groups to classify them in (irreducible) representations thereof, and the proposal that quarks are the constituents of hadrons, which are the particles capable of interacting through strong interactions.

Yang and Mills \cite{YangMills1954} proposed the extension of gauge theories to non-Abelian groups in order to describe the weak and strong nuclear forces. Finally, the works of Glashow \cite{Glashow1961}, Weinberg \cite{Weinberg1967} and Salam \cite{Salam1968} led to the formulation of the Standard Model, based on the gauge group $SU(3) \times SU(2) \times U(1)$, encompassing three of the four known fundamental forces: electromagnetic force (QED, mediated by massless photons), weak force (mediated by massive W and Z bosons) and strong force (QCD, mediated by massless gluons). 

Again, to ascertain the consistency of the model, it was necessary to study whether it was renormalizable, {\emph i.e.} to tackle the problem of infinities in non-Abelian gauge theories calculations. This was accomplished by `t Hooft and Veltman \cite{tHooft:1972tcz}, while Gross and Wilczeck \cite{GrossWilczek1973} and Politzer \cite{Politzer1973}discovering the asymptotic freedom of quantum chromodynamics (QCD) were finally able to clarify the puzzling properties of strong interactions.

In this short summary, we repeatedly mentioned the difficulties of quantum field theories in dealing with infinite degrees of freedom in perturbative calculations. Lattice gauge theory (LGT) was proposed as a powerful non-perturbative approach to studying quantum field theories, particularly quantum chromodynamics (QCD). Following the seminal papers of K. Wilson in 1974 \cite{Wilson1974}, it consisted in discretizing the spacetime onto a lattice, allowing for numerical calculations of observables in strongly coupling, where perturbative expansion fails (see, for instance, \cite{Creutz1983} and references therein). Gauge fields are represented as links between lattice sites in a way that preserves gauge invariance.

Kogut and Susskind (KS) \cite{Kogut_Susskind_75_Hamiltformul, Kogut_intro} proposed one of the first formulations of LGT for discrete and continuous (compact) groups such as $\mathbb{Z}_2$, $D_3$, $U(N)$ and $SU(N)$ (see \cite{Zohar_2015_formlgtquantumsimul} for a complete overview of the Hamiltonian formulation of this model).
Recent advances in algorithms and computing power have significantly improved the precision of lattice calculations \cite{Martinez_2016}, making them an essential tool in theoretical physics.

In particular, LGT is an essential tool for investigating phase transitions in spin models \cite{Fradkin_1978_Ordergaugetheory,Fradkin_1980_fermionreplattice} and the emergence of gauge bosons from quantum order of vacuum \cite{Wen_2005_introquantumorder,Levin_2006_Quantumether}.
The engineering of digital simulations of lattice gauge theories \cite{Zhou_2022,Surace_2020,Zache_2018,Wiese_2013,Irmejs_2023,Zohar_2013_quantumsimgauge,Zohar_2012_simulcompactquantum,Popov_2024_variatquantumsim,Brennen_2009_simquantumdouble,Banuls_2020,Tagliacozzo_2013_opticalabelianlattice,preskill_2018,falcao_2024}, allows us to study totally unexplored regimes in the context of Standard Model \cite{Bauer_2023}, nuclear interactions \cite{Robin_2024_magicnuclear} and quantum gravity \cite{Cepollaro_2024_SEQT,polino2022photonic,Czelusta_2021}.

In recent years, LGT has attracted increasing attention
because the amount of resources needed for a simulation on the lattice is
drastically reduced as local Hilbert spaces shrink from being infinite-dimensional to
being finite-dimensional.

The objective of this paper is to study the {\itshape non-stabilizer} properties of the Hilbert spaces in lattice gauge theories. Non-stabilizerness is a property of quantum states related to the simulability of a theory: it was proven by Gottesman \cite{gottesman1998_heisenberg} that there exists a set of states and operations, called stabilizer states and Clifford operations, which can be efficiently simulated by a classical computer, whereas non-stabilizer states and non-Clifford operations require exponential time for simulation, necessitating the speed-up provided by a quantum computer.  We focus on the amount of non-Clifford resources required to enforce the gauge constraint on a lattice with a discrete Abelian group. In order to {\itshape quantify} these resources we use Stabilizer Entropies \cite{Leone_2022_SRE,Wang2023_SEqudit} as measures of non-stabilizerness. 
Stabilizer Entropies have proven to be an important tool in the study of  quantum chaos \cite{Leone_2021,Oliviero_2021}, in decoding algorithms from black hole's Hawking radiation \cite{Leone_2022,oliviero2024unscramblingquantuminformationclifford} and in the holographic framework of AdS/CFT correspondence \cite{White_2021,Cao_2024}.

In simulations of gauge theories, one typically works with fundamental degrees of freedom. Gauge invariance is then imposed by constructing gauge-invariant states via quantum circuits. We investigate the average cost of this operation in terms of non-Clifford resources and point out the case in which using fundamental degrees of freedom has no cost with respect to the gauge-invariant ones.

The main result of the paper is that lattice gauge theories with a discrete abelian group do not require resources beyond Clifford operations for simulations and can therefore be efficiently reproduced by a classical computer. In contrast, we argue that non-abelian gauge theories should be more complex to simulate, and provide an example that demonstrates this increased complexity.

This article is organized as follows: in \cref{sec:SE} we give a short review on the stabilizer formalism and define the measures of stabilizer entropy and stabilizer entropy gap that define the resources needed in the projection; in \cref{sec:LGT} we provide the setting of lattice gauge theory, with focus on the construction of gauge invariant Hilbert space via the action of local projectors; \cref{sec:Z2} and \cref{sec:Zl} represent the main result of the paper, that is the proof, via direct calculation, that the Stabilizer Entropy Gap is 0 for gauge theory with discrete abelian group; in \cref{sec:SU(2)} we discuss an existing example in the literature and the substantial differences between abelian and non-abelian gauge symmetries in the framework of simulations, based on the results of the previous sections. 

We conclude the article by summing up the results and by discussing the future perspectives of this work, in particular towards the direction of quantifying the amount of computational resources needed to efficiently simulate fundamental physics in a lab.

\section{Stabilizer Entropy}\label{sec:SE}
In this section, we review the stabilizer formalism and some recent results and technical tools that will be useful in the following sections.
\subsection{Stabilizer Entropy}
    Consider $\hi\simeq\mathbb{C}^{2\ot n}$ the Hilbert space of a $n$-qubit system and its Pauli group $\mathbb{P}_n=\pauli_1^{\ot n}$ with $\mathbb{P}_1=\{\Id,X,Y,Z\}$ the Pauli group of a single qubit. The \textit{Clifford group} $\mathcal{C}(n)\subset \mathcal{U}(n)$ is defined as the normalizer of the Pauli group \cite{gottesman1998_heisenberg}, namely, $\mathcal{C}(n):=\{ C\in \mathcal{U}(n)\,,\, \text{s.t. }\,\forall P\in \pauli_n\,,CPC^{\dag}=P'\in \pauli_n\}$; in short, $\mathcal{C}(n)$ is the group of unitary operators that transform a Pauli string into an other. 
    
    We define the \textit{computational basis} $\{\ket{i}\}_{i=1}^{d=2^n}$ of $\hi$ as the common eigenbasis of the operators $\{\Id,Z\}^{\otimes n}$, and the set PSTAB of pure \textit{stabilizer states} as the Clifford orbit of the computational basis:
\begin{equation}
        \stab=\{C\ket{i}\,,C\in \mathcal{C}(n)\}\,.
    \end{equation}
   Stabilizer states and Clifford operations play a central role in quantum computation: the Gottesman-Knill theorem \cite{gottesman1998_heisenberg} states that any quantum process represented by Clifford operations acting on initial stabilizer states can be simulated by a classical computer in polynomial time.
   This implies that one needs states and gates beyond the stabilizer formalism to obtain quantum advantage in computation: the resource needed is referred to as \textit{non-stabilizerness} or \textit{magic}.
   
   Recently, $\alpha$-Stabilizer R\'enyi Entropy \cite{Leone_2022_SRE} has been proposed as an entropic measure of non-stabilizerness.
   
   Define the probability distribution associated with the tomography of a $n$-qubit pure state $\psi=\ketbra{\psi}$, namely $\Xi_P(\ket{\psi}):=d^{-1}\tr^2(P\ketbra{\psi})$, with $P$ being a Pauli operator.
   The $\alpha$-Stabilizer R\'enyi Entropy (SE) is 
\be M_{\alpha}(\psi):=(1-\alpha)^{-1}\log\sum_P\Xi_P^{\alpha}(\ket{\psi})-\log d
\ee
    while its linear counterpart, the linear stabilizer entropy (LSE) is 
    \ba  \mlin(\psi):&=&1-d\norm{\Xi_P(\ket{\psi})}^2 _2 \nonumber \\
            &=&1-d^{-1}\sum_{P\in\pauli{n}}\tr^4(P\ketbra{\psi}) \ea
Both measures are: (i) faithful, i.e. $M(\psi)=0 \Leftrightarrow \psi \in \stab$, otherwise $M(\psi)>0$; (ii) invariant under Clifford operators, namely $M(C\psi C^\dag)=M(\psi)$.

These measures quantify the non-stabilizerness of qubit states by the entropy of the state tomography: states with high values of SE require an exponential amount of classical resources to be simulated. 
Unlike other measures of non-stabilizerness (see for example \cite{Wang_2019,Veitch_2014}), SE  and LSE do not require any minimization processes and can be both computed and measured in experiments \cite{Oliviero2022}.
In this work we will focus on $\mlin$ due to our interest in Haar and Clifford averages. The linearity of $\mlin$ makes it particularly suitable to average techniques discussed in detail in \cite{Leone_2021_QCIQ}. 
Moreover, $\alpha$-SE (for $\alpha\geq2$) and LSE have recently been validated as \textit{magic monotones} in multi-qubit systems \cite{Leone_2024_stabentmonotone}. These measures serve as effective measures for non-stabilizer resources in various settings, highlighting their utility in the analysis of quantum systems.

\subsection{Stabilizer Entropy gap}
Our goal is to quantify the resources required to obtain a gauge-invariant Hilbert space. Typically, gauge invariance is enforced by projecting onto the subspace of states that are invariant under the action of a gauge group. To address this, we first establish a general framework for measuring the resources associated with any projection map, not limited to those associated with gauge invariance. Then, we will apply this framework to the specific case of gauge-invariant subspaces, considering both abelian and non-abelian symmetry groups.

Consider a Hilbert space $\hi$ and a subspace $\hi_{\Pi}=\Pi\hi$, where $\Pi$ is a projector, and $\hi=\hi_{\Pi}\oplus\hi_{\Pi}^{\perp}$. Let us define the two basis $\mathcal{B}_{\hi}=\{\ket{i}\}_{i=1}^d$ and $\mathcal{B}_{\hi_{\Pi}}=\{\ket{j}\}_{j=1}^{d_{\Pi}}$ with $d_{\Pi}={\rm rank}(\Pi)\leq d$. 
For prime dimension $l$, a natural generalization of the Pauli group is given by the Weyl-Heisenberg group \cite{gross2006hudson}
\be
\tilde{\mathcal{D}}^{(l)}_1=\left<\{D_{(p,q)}\}\right>=\{\tau^aD_{(p,q)}\}\quad \text{with}\quad a,p,q\in Z_l
\ee
i.e. the group generated by the $l^2$ operators defined as
\be \label{whgen}
D_{(p,q)}=\tau^{-pq}Z^{p}X^{q} \quad \text{with}\quad \tau=e^{\frac{i}{l}\pi} \quad p,q\in Z_d 
\ee
and 
\be
X\ket{j}=\ket{j\oplus_l 1} \quad
Z\ket{j}=\omega^j\ket{j} \quad \omega=e^{\frac{2\pi i}{l}}\,. \ee
The construction of Weyl-Heisenberg operators generalizes straightforwardly for $n$ qudit systems via tensor product:
\be
\tilde{\mathcal{D}}_n ^{(l)}=\tilde{\mathcal{D}}_1 ^{(l)}\otimes\underbrace{\dots}_n\otimes\tilde{\mathcal{D}}_1 ^{(l)} 
\ee
For even or odd-nonprime dimensions, the set generated by the operators in expression \eqref{whgen} are not guaranteed to have an inverse, so the set fails to be a group: this is due to the fact that, for $l$ prime, $\mathbb{Z}_l$ is an \textit{algebraic field}, so our qudits are actually described by a $l$-dimensional Hilbert space, which is not the case for non-prime $l$, as it is necessary for any vector space to be constructed upon a field. However, it is possible to obtain a description for generic number of levels by tensor product of Hilbert spaces of dimensions equal to the prime decomposition of the desired number, along with the corresponding Weyl-Heisenberg operators for each factor. Hence, for the remainder of this work we are going to stick with prime dimensions (2 included) without loss of generality.
The stabilizer entropies introduced in the previous subsection have an analogous form when the Pauli decomposition is substituted with the Weyl-Heisenberg decomposition \cite{Wang2023_SEqudit}.

To facilitate future discussions, we introduce the projector  \be Q=\frac{1}{d^2}\sum_{D_{(p,q)}\in \mathcal{D}^{(l)}_n}(D_{(p,q)}\otimes D_{(p,q)}^{\dag})^{\otimes 2} \ee 
with $d=l^n$.

This projector plays a crucial role in the subsequent analysis.
Indeed, we can use it to define a more concise expression for the Linear Stabilizer entropy:
\be\label{mlin} M_{\lin}(\psi)=1-d\tr(Q\psi^{\ot 4}) \ee 

In general, the projection onto a subspace $\hi_{\Pi}$ of a larger Hilbert space has a cost in terms of non-stabilizer resources. We are interested in quantifying the average amount of non-Clifford resources associated with the projection onto a {\itshape gauge-invariant} subspace. To this end, we use the \textit{Stabilizer entropy gap} (SE gap) \cite{Cepollaro_2024_SEQT}, defined as 
\begin{equation}
         \Delta M(\hi_G):= M_{\lin} ^0 -M_{\lin} ^G
         =\exv_{U_G} M_{\lin} ^0(\psi_{U_G})- \exv_{U_G}M_{\lin} ^G (\psi_{U_G})
\end{equation}
with $\exv_{U_G}$ denoting the Haar average over the unitary operators acting on the gauge invariant Hilbert space $\hi_G=\Pi_G\hi$, and $\psi_{U_G}=U_G \psi U_G ^\dag$, and
\begin{equation}
    M_{\lin} ^i 
    (\psi):=1-d_i \tr(Q^{i}\psi^{\ot 4})\,, i\in \{0,G\}
\end{equation}
denoting the non-stabilizer resources calculated respectively in the Weyl-Heisenberg structures of the original Hilbert space and in the gauge-invariant subspace. 

The SE gap measures the non-stabilizer resources required to project onto a subspace in a \textit{state-independent} manner. Given that the difference in SE is averaged over all possible gauge-invariant states, a nonzero value of the gap indicates an inherent incompatibility between the Weyl-Heisenberg operators structures of the ambient space and the projected gauge-invariant subspace.

Using the expression given by \cref{mlin}, we can rewrite the SE gap as
\begin{equation}
    \Delta M (\hi_G)=d_G \exv_{U_G} \tr(Q^G \psi_{U_G} ^{\ot 4})-d_0 \exv_{U_G} \tr(Q^0\psi_{U_G} ^{\ot 4})\,. 
\end{equation}
By applying Lemma 1 from \cite{Cepollaro_2024_SEQT} and the results derived for the Haar average over the unitary group \cite{Mele_2024, Roberts_Yoshida_2017}, we can use the proportionality between the average in the subspace and the one in the total space:
\be
        \exv_{U_G} \tr(Q^0\psi_{U_G} ^{\ot 4})=c(d_0,d_G)\exv_{U_0} \tr(Q^0\Pi_G ^{\ot 4}\psi_{U_0} ^{\ot 4})
\ee
with
\begin{equation}
    c(d_0,d_G)=\binom{d_0+3}{4}\binom{d_G+3}{4}^{-1}\,.
\end{equation}

For future convenience, we show the condition for a zero gap invariant subspace: simple manipulation of the previous expressions shows that
\begin{equation}\label{0gap}
        \Delta M(\hi_G)=0\iff \tr(Q^0 \Pi_{\sym} ^0 \Pi_G ^{\ot 4})=\frac{d_G}{d_0}\tr(Q^G \Pi_{\sym} ^G)
\end{equation}
With $\Pi_{\sym}^i\,,i\in \{0,G\}$ being the projector onto the $S_4$ permutationally invariant subspace respectively of $\hi_0 ^{\ot 4}$ and $\hi_G ^{\ot 4}$. A detailed discussion on the projector $Q$ can be found in \cite{Leone_2021_QCIQ}, and several applications of this projector are shown in details in \cref{appHW,appz2}.

\section{Short overview of Lattice Gauge Theory}\label{sec:LGT}
In this work, we use a model of Lattice Gauge Theory (LGT) inspired by the Kogut-Susskind formulation \cite{Kogut_Susskind_75_Hamiltformul}, where matter fields reside on the vertices of a square lattice and gauge fields mediate interactions along the links. However, since our primary focus is on a pure gauge theory, we concentrate on the behavior and properties of the links Hilbert space. The model we investigate draws inspiration from the Toric code proposed by Kitaev \cite{Kitaev_2003_faultolquantum}. Kitaev proposes a model in which the gauge-invariant subspace is characterized by projectors $\Pi_G$ that are written in terms of \textit{star operators} which strictly depend on the gauge group G. This approach allows us to construct the gauge-invariant Hilbert space as $\hi_G=\Pi_G\hil$, thereby providing a robust framework for exploring the average resources within a quantum gauge theory.

\subsection{Gauge invariant Hilbert space}
The initial element required to formulate a lattice gauge theory (LGT) is a symmetry group G. In this discussion, we choose to focus on the case of \textit{discrete groups}, whose finite nature permits a straightforward description of the Hilbert space associated with the lattice. This choice allows us to analyze the non-stabilizer resources that characterize a gauge invariant physical system within a very concise mathematical setting.

Each link is endowed with a local $l$-dimensional Hilbert space $\hil_l$ and the total Hilbert space is given by $\hi_0=\hil_l^{\ot N}$. The link Hilbert spaces are chosen to be isomorphic to the group algebra Hilbert space $\mathbb{C}(G)$ \footnote{Let G be a finite group. The group algebra $\mathbb{C}(G)$ is defined as the vector space of complex linear combinations of group elements. This space has the structure of a Hilbert space with orthonormal basis given by $\{\ket{g}\,\,\text{s.t.}\,\, g\in G\}.$ and scalar product $\braket{g|h}=\delta_{g,h}$.}:
\be g\in G\to \ket g_l \in \hil_l\quad \text{such that}\quad \braket{h|g}=\delta_{hg} \ee 

The main request of this approach is that the local Hilbert spaces have to be isomorphic to some representation of the group: this implies that, for discrete group $\text{dim}\hil_l=|G|$

Choosing the group algebra $\mathbb{C}(G)$ allows us to use the fundamental representation of the group, i.e. $R_f(G)$. By doing so, we can define the left action of an element $g\in G$ on the Hilbert space $\hil_l$ as $X(g)\in R_f(G)$ defined by:
\be X(g)\ket{h}=\ket{gh}\quad \implies \quad X(g)=\sum_{h\in G}\ket{gh}\bra{h} \ee

Gauge invariance is obtained by the local action on each site of the lattice with the action of the \textbf{star operator} $A_s(g)$ \cite{Kitaev_2003_faultolquantum,Hamma_2005_bipartite_entanglement,Hamma_2005_ground_state_entanglement}. 

Let us define the operator 
\ba 
A_s(g)&=&\bigotimes_{l\to s}X_l(g) \\ A_s(g)\ket{h_{l_1},h_{l_2},h_{l_3},h_{l_4}}&=& \ket{gh_{l_1},gh_{l_2},gh_{l_3},gh_{l_4}} \nonumber \ea 
where $s$ label the site, $\ket{h_{l_i}}\in\hil_{l_i}$ and $l\to s$ is the shorthand notation to indicate the link entering the site.
 
States in the Hilbert space are invariant under gauge transformations if they are eigenvectors of $A_s(g)$ with eigenvalues $+1$ for all group element:
\be A_s(g)\ket{\psi_{\text{phys}}}=\ket{\psi_{\text{phys}}} \quad \forall g\in G \ee 
The physical gauge invariant Hilbert space $\hil_G$ is the set of states in $\hi_0$ such that
\be \hil_G=\{\ket{\psi}\,\,\text{s.t.}\,\,\bigotimes_{s\in \mathcal{L}}A_s(g)\ket{\psi}=\ket{\psi},\,\forall g\in G\} \ee 

$\hil_G$ is a subspace of the total Hilbert space, obtained with projection on each site of the lattice where the projector is given by the sum of all the star operators acting on the link:
\be 
\Pi_G^{(s)}=\frac1{|G|}\sum_{g\in G}A_s(g) \ee 
and the projector onto the gauge invariant sector of the total Hilbert space of the lattice reads
\begin{equation}
    \Pi_G=\prod_{s\in \mathcal{L}}\Pi_G ^{(s)}\,.
\end{equation}

Now, consider $\mathfrak{G}:=\{A_s(g)\,,g\in G, s\in \mathcal{L}\}$ the set of all possible star operators and $\mathcal{G}:=\langle \mathfrak{G}\rangle$ the group generated by all possible products of elements i $\mathfrak{G}$. 
The projector onto $\hi_G$ can be written as \cite{Flammia_Hamma_Hughes_Wen_2009}
\begin{equation}\label{pitot}
    \Pi_G=\frac{1}{|\mathcal{G}|}\sum_{h\in \mathcal{G}}h\,,
\end{equation}
and it fully characterizes the gauge invariant system described by the graph, as it contains all the information on the local gauge group. 

\subsection{Lattice orientation and non-abelian gauge groups}
Consider a lattice $\mathcal{L}$ composed of sites $s$ and links $l$. Typically, it is essential to work with oriented graphs; this necessity arises from the physical interpretation of a link as a gauge current moving between sites. In this framework, the gauge invariant condition, enforced by the Gauss constraints at each site, ensures current conservation. Although the orientation of the lattice is not particularly important for abelian groups, it becomes a necessary technical tool for non-abelian gauge simmetry when it comes to the definition of local operators defined on the links. Nontheless, it is worth noticing that lattice orientation is a convention that bears no physical significance as long as it is coherently applied throughout the calculations. States in the Hilbert space should indeed incorporate this orientation information, as the change in orientation corresponds to the inversion of the current: \be \ket h_l \to \ket{h^{-1}}_l \ee
For non-abelian groups, altering the orientation of a link also modifies the group action, represented by the local adjoint operator $X(g)^{\dag}$. 
However, in the context of our current analysis, the orientation of the lattice does not significantly influence the results; since gauge invariance is achieved by summing the star operators over all group elements , changing the orientation does not influence the structure of the sum, hence leaving the gauge-invariant space invariant. The orientation would become particularly relevant when gauge-invariant operators, such as the plaquette operator or a Hamiltonian operator, are studied. Although this is not the main focus of this work, these considerations will be important in future developments of our research.

\section{SE gap of the $Z_2$ lattice gauge theory}\label{sec:Z2}

Let us now look at the simplest example of LGT with gauge group $G=Z_2=\{e,g\}$. $Z_2$ is the smallest abelian group with the following multiplication table:
\be e\cdot e=e\quad e\cdot g=g \quad g\cdot g =e \ee 
Since the group order is 2, we only need one qubit Hilbert space $\mathbb{C}^2$ for each link of the graph to construct the Hilbert space of the whole lattice. A square lattice $\mathcal{L}$ with $L$ links and periodic boundary conditions is equipped with the Hilbert space $\hi_0=\mathbb{C}^{2\otimes L}$. The single link group algebra Hilbert space is $\hil_l=\text{span}\{\ket {0=e},\ket{1=g}\}$.

The action of $Z_2$ on the link states is given by the left action, expressed in the fundamental representation as $X(g)=\sum_{h\in G}\ket{gh}\bra{h}$. 

The two independent actions are: 
\ba 
X(e)&=&\ket 0\bra 0 + \ket 1 \bra 1 \nonumber \\
X(g)&=& \ket 0 \bra 1 + \ket 1 \bra 0 \ea 
The matrix representation of these operators in $\text{Lin}(\hil_l)$ are respectively the identity matrix and the $X$ Pauli matrix:
\be X(e)=\Id_2\quad X(g)=X \ee 
For each site, comprised of links $\{l_1^{(s)},l_2^{(s)},l_3^{(s)},l_4^{(s)}\}$, the star operators can be written as
\ba A_s(e)&=&\Id_{l_1^{(s)}}\ot \Id_{l_2^{(s)}}\ot \Id_{l_3^{(s)}} \ot \Id_{l_4^{(s)}} \nonumber \\  
A_s(g)&=& X_{l_1^{(s)}} \ot X_{l_2^{(s)}}\ot X_{l_3^{(s)}} \ot X_{l_4^{(s)}} \ea 

On a single site $s$, the form of the gauge-invariant projector is $\Pi_s=\frac{\Id^{\ot 4}+X^{\ot 4}}{2}$ and the projector onto the whole lattice, according to Eq. \eqref{pitot}, reads
\begin{equation}
    \Pi_G=\frac{1}{|\mathcal{G}|}\sum_{P\in \mathcal{G}}P\,.
\end{equation}
Note that $\mathcal{G}:=\langle\{A_s(g)\,,g\in Z_2, s\in \mathcal{L}\}\rangle$ is a subgroup of \textit{commuting} Pauli operators. The dimension of the gauge-invariant Hilbert space is $d_G=d_0/\cg$.
Once the whole structure for both total and gauge-invariant Hilbert spaces is defined, we can calculate the magic gap between the two systems: in short, we compute the difference of the average linear stabilizer entropies (of the gauge-invariant Hilbert space) calculated respectively with respect to the Pauli operators in $\hi_0$ and $\hi_G$.

After a long calculation reported in detail in \cref{appz2}, we find the following results:
\ba \label{spz20}
   \tr(Q^0 \Pi_{\sym} ^0 \Pi_G ^{\ot 4})&=&\frac{1}{6\cg}\left(\frac{d_0 ^2}{\cg ^2}+3\frac{d_0}{\cg}+2\right)\\
        \tr(Q^G \Pi_{\sym} ^G)&=&\frac{1}{6}(d_G^2+3d_G+2)
   \ea 
Looking at \cref{0gap} and at the relation between the dimensions of the two Hilbert space, we conclude that the zero-gap conditions is satisfied: \be \Delta M(\hi_{Z_2})=0 \ee 
This means that the Pauli structure of the gauge-invariant Hilbert space can be faithfully reproduced by the Pauli structure of the total Hilbert space with no cost in terms of nonstabilizer resources. A non-vanishing gap would imply that a speed-up provided by a quantum processor should be necessary to efficiently simulate any process of the theory starting from the total Hilbert space.

\section{LSE GAP of the $Z_l$ lattice gauge theory} \label{sec:Zl}
In this section we generalize the previous result, extending our consideration to the group $Z_l=\{0,g_1,\ldots,g_{l-1}\}$, $\absval{G}=l$.
Each link of the lattice $L$ is now equipped with a qudit Hilbert space $\hil_l=\mathbb{C}^l=\text{span}\{\ket{0=e_G},\ket{1=g_1},\ldots \ket{l-1=g_{l-1}}$.

At this point, in order to study the non-stabilizer properties of this type of system, we need to restrict our analysis to prime $l$: by doing so, the Weyl-Heisenberg group is well defined \cite{gross2006hudson} and we can use the generalization of stabilizer entropies for a qudit Hilbert space. 

Given the group algebra basis, we can write the representation of group elements $\{e,g,\ldots,g_{l-1}\}$ as powers of the Weyl-Heisenberg operator $\operatorname{X}$
\ba 
X(e)&=&\ket 0\bra 0+ \ldots \ket{l-1}\bra{l-1}=\Id_l  \\
X(g_1)&=&\ket 0 \bra 1+ \ldots \ket{l-1}\bra 1 =\operatorname{X} \\
\ldots \\
X(g_{l-1})&=&\ket 0 \bra{d-1}  + \ldots \ket{l-1}\bra{l-2}=\operatorname{X}^{l-1}
\ea 
Accordingly, the star operators on a single site s of the square lattice read
\ba  
A_s(e)&=&\Id_{l_1^{(s)}}\ot \Id_{l_2^{(s)}}\ot \Id_{l_3^{(s)}} \ot \Id_{l_4^{(s)}} \\
A_s(g_1)&=& X_{l_1^{(s)}} \ot X_{l_2^{(s)}}\ot X_{l_3^{(s)}} \ot X_{l_4^{(s)}}\\
\ldots \\
A_s(g_{l-1})&=& X_{l_1^{(s)}}^{l-1} \ot X_{l_2^{(s)}}^{l-1}\ot X_{l_3^{(s)}}^{l-1} \ot X_{l_4^{(s)}}^{l-1}
\ea 
Therefore, we can define the group generated by the star operators at all sites of the graph as a subgroup of commuting Heisenberg Weyl operators $\mathcal{G}:=\langle\{A_s(g)\,,g\in Z_l, s\in \mathcal{L}\}\rangle$. By doing so, the projector into the gauge invariant Hilbert space reads
\begin{equation}
    \Pi_G=\frac{1}{|\mathcal{G}|}\sum_{D\in \mathcal{G}}D\,.
\end{equation}
To calculate the magic gap we need to compute the terms $\tr Q^0\Pi_{\sym}^0\Pi_G^{\ot 4}$ and $\tr Q^G\Pi_{\sym}^G $. Direct calculation (see \ref{appzl} for the details) leads to:
\ba \tr  Q^0\Pi_{\sym}^0\Pi_G^{\ot 4} &=&\frac{1}{24}\left[\frac{3d_0^2}{\absval{G}^3}+\frac{12 d_0}{\absval{G}^2}+\frac{9}{\absval{G}}\right] \\
\tr Q^G\Pi_{\sym}^G &=&\frac{1}{24}\bigg[3d_G^2+(1-\text{mod}_2(l))(4^{n}-1) \nonumber \\  &+&12d_G+9\bigg] \ea
Notice that, in this calculation, we did not restrict to the case of odd dimensions, in order to prove, as a consistency check, that in the limit $l\to 2$, the formalism was consistent with the Pauli case of $Z_2$. 

We conclude this analysis on abelian groups, by saying that the magic gap of $\mathbb{Z}_l$ lattice gauge theory is:
\be
\Delta M(\hil_{Z_l}) =d_G\exv_{U_G}\tr Q_G\psi_{U_G}^{\ot 4}-d_0\exv_{U_G}\tr Q^0\psi_{U_G}^{\ot 4}=0
\ee

\section{LSE gap for a non-abelian group: SU(2)}\label{sec:SU(2)}
We proceed with our discussion by analyzing a significant difference between what we obtained with abelian groups and an example coming from \cite{Cepollaro_2024_SEQT}, concerning the $SU(2)$-gauge invariant Hilbert space on the site of a lattice. 

Consider an open lattice composed of one site with four links. At each link, we attach a Hilbert space associated to an irreducible representation of $SU(2)$: in particular, choosing the fundamental (spin $\frac 12)$ representation leads to a qubit Hilbert space $\hi_{\frac 12}$ on each link, similar to the case of the $Z_2$ gauge theory. 

The non-gauge invariant Hilbert space is given by the tensor product of the Hilbert space of the four links, that is, $\hi_{\frac 12}^{\ot 4}$. Imposing gauge invariance requires Peter-Weyl decomposition of this tensor product structure, which reads:
\be 
\hil_{\frac 12}^{\ot 4}=2\hi_0\oplus 3\hi_1\oplus \hi_2 \ee 
which corresponds to two singlets, three triplets and one quintuplet Hilbert spaces. 
The gauge invariant subspace is obtained via group averaging and corresponds to the singlet Hilbert space $2\hi_0$, which is a $2$-dimensional Hilbert space. This means that the rank-$2$ projector $\Pi_{SU(2)}$ that realizes $2\hi_0=\Pi_{SU(2)}\hi_{\frac 12}^{\ot 4}$ can be used to calculate the magic gap, with the exact same procedure as we showed in \ref{appz2} and \ref{appzl}.
In \cite{Cepollaro_2024_SEQT}, it was shown that this system with non-abelian gauge symmetry exhibits a non-vanishing magic gap. 

In particular, for single site lattice, the LSE gap is
\be \Delta M(\hi_{SU(2)})=8/45 \,,\ee
which means that $SU(2)$ gauge invariance needs nonstabilizer resources to be implemented.

\section{Conclusions}

Discrete abelian gauge theories seem to not require non-stabilizer resources for their simulation. This would mean that enforcing the gauge constraint on a lattice does not require computational costs, enabling efficient simulation of the theory by using only classical resources, by focusing on the non gauge-invariant degrees of freedom. 

On the other hand, the same efficiency doesn't seem to be achievable for non-abelian gauge theories. In this paper we shortly discussed a non-abelian counterexample ($SU(2)$ fundamental representation) where the resource cost is non-zero, indicating a significant difference in the computational requirements. This distinction suggests that non-abelian gauge theories inherently demand more computational resources.

These observations may point towards a broader characterization of gauge theories based on the interplay between simulability and non-commutativity. We expect non-abelianity of the gauge group directly affects the computational cost in terms of non-stabilizer resource. This effect is expected to mirror the way in which non-abelian structures influence certain phenomena associated with another resource, namely entanglement, i.e. asymmetric behavior in the Page curve in many body systems with $SU(2)$ symmetry \cite{bianchi_2024}.
 Further investigation would focus on formally proving this intuition and pointing out the feature of non-abelian theory increasing its computational cost by looking at the interplay between the Clifford group and continuous non-abelian symmetry groups \cite{Mitsuhashi_2023}. This potential characterization could provide valuable insight into the resource demands of different gauge theories and guide the development of more efficient simulation algorithms, as well as a more efficient resource use for experimental implementations of said LGTs, as it is known that magic also induces upper bounds on the maximum achievable fidelity of experimental realizations of quantum states \cite{Cepollaro_2024_SEQT,Leone_nonstab_hardness_direct_fidelity}.

\section*{Acknowledgements}
AH acknowledges support from PNRR MUR project PE0000023-
NQSTI and PNRR MUR project CN 00000013 -ICSC.

\bibliographystyle{unsrtnat}
\bibliography{ref_lgt}
\ignore{

}
\newpage
\appendix

\section{Calculation of the $Z_2$ SE gap}\label{appz2}
In this section, we show the explicit calculations necessary to prove that the $Z_2$ has zero SE gap.
Let us start by the definition of the projector $\Pi_{\sym}$:
\begin{equation}
    \Pi_{\sym}:=\frac{1}{24}\sum_{\pi \in S_4}\sswap_{\pi}\,,
\end{equation}
where $\sswap_\pi$ are the representation of the permutation group of four objects acting on four copies of a Hilbert space $\hi$. Given any basis $\{\ket i\}$, they read 
\begin{equation}
   \sswap_{\pi}:=\sum_{i,j,k,l}\ket{\pi(ijkl)}\bra{ijkl}\,.
\end{equation}
Recalling the result in \cite{Leone_2021_QCIQ}
\begin{equation}\label{trqpisym}
    \tr(Q\Pi_{\sym})=\frac{1}{6}(d^2+3d+2)\,,
\end{equation}
we now show the calculations leading to Eq. \eqref{spz20}: our object of interest is 
\begin{equation}\label{sp0}
    {\rm SP}_0(\hi_G):=\tr(Q^0 \Pi_{\sym} ^0 \Pi_G ^{\ot 4})\,.
\end{equation}
In order to complete this calculation, let us write the expression of $\Pi_G ^{\ot 4}$ explicitly:
\begin{equation}
    \Pi_G ^{\ot 4}=\frac{1}{\cg^4}\sum_{P_1, P_2, P_3,P_4 \in \mg}P_1\ot P_2 \ot P_3\ot P_4\,,
\end{equation}
and also write the operator $Q^0 \Pi_{\sym} ^0$ in Pauli operators, since they provide a (orthogonal) basis for the space of linear operators acting on $\hil$:
\begin{equation}
    \begin{split}
        Q^{0}\Pi_{\sym}=\frac{1}{d_0 ^4}\sum_{Q_1,Q_2,Q_3,Q_4\in \pauli_L}\tr[Q^{0}\Pi_{\sym} ^0 (Q_1\ot Q_2\ot Q_3\ot Q_4)]Q_1\ot Q_2\ot Q_3\ot Q_4\,.
    \end{split}
\end{equation}

Substituting these expressions in Eq. \eqref{sp0}, we get
\begin{equation}\label{sp0bfperm}
    \begin{split}
        {\rm SP}_0(\hi_G)&=\frac{1}{d_0 ^4\cg^4}\sum_{P_1, P_2, P_3,P_4 \in \mg}\sum_{Q_1,Q_2,Q_3,Q_4\in \pauli_L}\tr[Q^{0}\Pi_{\sym} ^0 (Q_1\ot Q_2\ot Q_3\ot Q_4)]\\&\tr(Q_1 P_1)\tr(Q_2 P_2)\tr(Q_3 P_3)\tr(Q_4 P_4)\\
        &=\frac{1}{\cg^4}\sum_{P_1, P_2, P_3,P_4 \in \mg}\sum_{Q_1,Q_2,Q_3,Q_4\in \pauli_L}\tr[Q^{0}\Pi_{\sym} ^0 (Q_1\ot Q_2\ot Q_3\ot Q_4)]\\
        &\delta_{Q_1,P_1} \delta_{Q_2,P_2} \delta_{Q_3,P_3} \delta_{Q_4,P_4}\\
        &=\frac{1}{\cg^4}\sum_{P_1, P_2, P_3,P_4 \in \mg}\tr[Q^{0}\Pi_{\sym} ^0 (P_1\ot P_2\ot P_3\ot P_4)]\,,
    \end{split}
\end{equation}
where we used the orthogonality relations of Pauli operators: $\tr(PP')=d_0\delta_{P,P'}$. 
Now what remains is to determine the coefficients of the Pauli decomposition of $Q\Pi_{\sym}$: writing $\Pi_{\sym}$ explicitly, it means we need to calculate objects of the form $\tr[(P_1\ot P_2 \ot P_3 \ot P_4)Q\sswap_{\pi}]$ for each permutation $\pi \in S_4$. Luckily, thanks to the $Q-\sswap$ equalities \cite{Leone_2021_QCIQ} we need only calculate six of these traces. The statement is that for every operator $\mathcal{O}\in \mathcal{B}(\hi_0 ^{\ot 4})$, the following equalities hold:
\begin{equation}\label{permoeq}
    \begin{split}
\tr(\mathcal{O}Q^0)&=\tr(\mathcal{O}Q^0\sswap^0 _{(12)(34)})=\tr(\mathcal{O}Q^0\sswap^0 _{(13)(24)})=\tr(\mathcal{O}Q^0\sswap^0 _{(14)(23)})\\
\tr(\mathcal{O}Q^0\sswap^0 _{(24)})&=\tr(\mathcal{O}Q^0\sswap^0 _{(1432)})=\tr(\mathcal{O}Q^0\sswap^0 _{(13)})=\tr(\mathcal{O}Q^0\sswap^0 _{(1234)})\\
\tr(\mathcal{O}Q^0\sswap^0 _{(14)})&=\tr(\mathcal{O}Q^0\sswap^0 _{(23)})=\tr(\mathcal{O}Q^0\sswap^0 _{(1342)})=\tr(\mathcal{O}Q^0\sswap^0 _{(1243)})\\
\tr(\mathcal{O}Q^0\sswap^0 _{(12)})&=\tr(\mathcal{O}Q^0\sswap^0 _{(34)})=\tr(\mathcal{O}Q^0\sswap^0 _{(1324)})=\tr(\mathcal{O}Q^0\sswap^0 _{(1423)})\\
\tr(\mathcal{O}Q^0\sswap^0 _{(132)})&=\tr(\mathcal{O}Q^0\sswap^0 _{(234)})=\tr(\mathcal{O}Q^0\sswap^0 _{(124)})=\tr(\mathcal{O}Q^0\sswap^0 _{(143)})\\
\tr(\mathcal{O}Q^0\sswap^0 _{(123)})&=\tr(\mathcal{O}Q^0\sswap^0 _{(142)})=\tr(\mathcal{O}Q^0\sswap^0 _{(243)})=\tr(\mathcal{O}Q^0\sswap^0 _{(134)})\,.
\end{split}
\end{equation}
This means we only need to calculate six traces:
\begin{equation}\label{trqpipauli}
    \begin{split}
        \bullet &({\rm id})\\
        &\tr[(P_1 \ot P_2 \ot P_3 \ot P_4) Q^0 ]=\frac{1}{d_0 ^2}\sum_P \tr(P P_1)\tr(P P_2)\tr(P P_3)\tr(P P_4)\\
        &=d_0 ^2\sum_P \delta_{P,P_1}\delta_{P,P_2}\delta_{P,P_3}\delta_{P,P_4}=d_0 ^2\delta(P_1=P_2=P_3=P_4)\\
        \bullet &(24)\\
        &\tr[(P_1 \ot P_2 \ot P_3 \ot P_4) Q^0  \ttt_{(24)}]=\frac{1}{d_0 ^2}\sum_P \tr[(P_1 P\ot P_2 P\ot P_3 P\ot P_4 P)\ttt_{(24)}]\\
        &=\frac{1}{d_0 ^2}\sum_P \tr(P_1 P)\tr(P_3 P)\tr(P_4 P P_2 P)=\sum_P \delta_{P_1 P}\delta_{P_3 P}\tr(P_4 P P_2 P)\\
        &=\delta_{P_1, P_3}\tr(P_4 P_1 P_2 P_1)=d_0\delta_{P_1, P_3}\delta_{P_4,P_1 P_2 P_1}\\
        \bullet &(14)\\
        &\tr[(P_1 \ot P_2 \ot P_3 \ot P_4) Q^0  \ttt_{(14)}]=\frac{1}{d_0 ^2}\sum_P \tr[(P_1 P\ot P_2 P\ot P_3 P\ot P_4 P)\ttt_{(14)}]\\
        &=\frac{1}{d_0 ^2}\sum_P\tr(P_2 P)\tr(P_3 P)\tr(P_4 P P_1 P)=\sum_P \delta_{P_2,P}\delta_{P_3,P}\tr(P_4 P P_1 P)\\
       &=\delta_{P_2,P_3}\tr(P_4 P_2 P_1 P_2)=d_0\delta_{P_2,P_3}\delta_{P_4 ,P_2 P_1 P_2} 
    \end{split}
\end{equation}
\begin{equation*}
    \begin{split}
        \bullet &(12)\\
         &\tr[(P_1 \ot P_2 \ot P_3 \ot P_4) Q^0  \ttt_{(12)}]=\frac{1}{d_0 ^2}\sum_P \tr[(P_1 P\ot P_2 P\ot P_3 P\ot P_4 P)\ttt_{(12)}]\\
         &=\frac{1}{d_0 ^2}\sum_P \tr(P_4 P)\tr(P_3 P)\tr(P_2 P P_1 P)=\sum_P \delta_{P_4,P}\delta_{P_3,P}\tr(P_2 P P_1 P)\\
         &=\delta_{P_3,P_4}\tr(P_2 P_4 P_1 P_4)=d_0\delta_{P_3,P_4}\delta_{P_2 ,P_4 P_1 P_4}\\
         \bullet &(132)\\
          &\tr[(P_1 \ot P_2 \ot P_3 \ot P_4) Q^0  \ttt_{(132)}]=\frac{1}{d_0 ^2}\sum_P \tr[(P_1 P\ot P_2 P\ot P_3 P\ot P_4 P)\ttt_{(132)}]\\
          &=\frac{1}{d_0 ^2}\sum_P \tr(P_4 P)\tr(P_2 P P_3 P P_1 P)=\frac{1}{d_0}\tr(P_2 P_4 P_3 P_4 P_1 P_4 )\\
          &=\delta_{P_2 P_4 P_3, P_4 P_1 P_4}\\
          \bullet &(123)\\
          &\tr[(P_1 \ot P_2 \ot P_3 \ot P_4) Q^0 \ttt_{(123)}]=\frac{1}{d_0 ^2}\sum_P \tr[(P_1 P\ot P_2 P\ot P_3 P\ot P_4 P)\ttt_{(123)}]\\
           &=\frac{1}{d_0 ^2}\sum_P \tr(P_4 P)\tr(P_3 P P_1 P P_2 P)=\frac{1}{d_0}\tr(P_3 P_4 P_1 P_4 P_3 P_4 )\\
           &=\delta_{P_3 P_4 P_1, P_4 P_2 P_4 }\,.
    \end{split}
\end{equation*}
Putting the pieces together, we get
\begin{equation}\label{trqperm}
    \begin{split}
        \tr[Q^0 \Pi_{\sym} ^0&(P_1\ot P_2 \ot P_3 \ot P_4)]=\frac{1}{6}\Big(d_0 ^2\delta_{(P_1=P_2=P_3=P_4)}+d_0\delta_{P_1, P_3}\delta_{P_4,P_1 P_2 P_1}\\
        &+d_0\delta_{P_2,P_3}\delta_{P_4 ,P_2 P_1 P_2}+d_0\delta_{P_3,P_4}\delta_{P_2 ,P_4 P_1 P_4}+\delta_{P_2 P_4 P_3, P_4 P_1 P_4}+\delta_{P_3 P_4 P_1, P_4 P_2 P_4 }\Big)
    \end{split}
\end{equation}
Substituting this expression in Eq. \eqref{sp0bfperm}, we get
\begin{equation}
    \begin{split}
         {\rm SP}_0(\hi_G)&=\frac{1}{6\cg^4}\sum_{P_1, P_2, P_3,P_4 \in \mg}\Big(d_0 ^2\delta_{(P_1=P_2=P_3=P_4)}+d_0\delta_{P_1, P_3}\delta_{P_4,P_1 P_2 P_1}\\
        &+d_0\delta_{P_2,P_3}\delta_{P_4 ,P_2 P_1 P_2}+d_0\delta_{P_3,P_4}\delta_{P_2 ,P_4 P_1 P_4}+\delta_{P_2 P_4 P_3, P_4 P_1 P_4}+\delta_{P_3 P_4 P_1, P_4 P_2 P_4 }\Big)\\
        &=SP_1+SP_2+SP_3+SP_4+SP_5+SP_6\,,
    \end{split}
\end{equation}
with
\begin{equation}
    \begin{split}
        {\rm SP}_1&:=\frac{d_0 ^2}{6\cg^4}\sum_{P_1, P_2, P_3,P_4 \in \mg}\delta_{(P_1=P_2=P_3=P_4)}\\
        {\rm SP}_2&:=\frac{d_0}{6\cg^4}\sum_{P_1, P_2, P_3,P_4 \in \mg}\delta_{P_1, P_3}\delta_{P_4,P_1 P_2 P_1}\\
        {\rm SP}_3&:=\frac{d_0}{6\cg^4}\sum_{P_1, P_2, P_3,P_4 \in \mg}\delta_{P_2,P_3}\delta_{P_4 ,P_2 P_1 P_2}
    \end{split}
\end{equation}
\begin{equation*}
    \begin{split}
        {\rm SP}_4&:=\frac{d_0}{6\cg^4}\sum_{P_1, P_2, P_3,P_4 \in \mg}\delta_{P_3,P_4}\delta_{P_2 ,P_4 P_1 P_4}\\
        {\rm SP}_5&:=\frac{1}{6\cg^4}\sum_{P_1, P_2, P_3,P_4 \in \mg}\delta_{P_2 P_4 P_3, P_4 P_1 P_4}\\
        {\rm SP}_6&:=\frac{1}{6\cg^4}\sum_{P_1, P_2, P_3,P_4 \in \mg}\delta_{P_3 P_4 P_1, P_4 P_2 P_4 }\,.
    \end{split}
\end{equation*}
We are going to calculate each piece separately: first one yields
\begin{equation}
         {\rm SP}_1=\frac{d_0 ^2}{6\cg^4}\sum_{P_1 \in \mg}1=\frac{d_0 ^2}{6\cg^3}\,,
\end{equation}
whereas
\begin{equation}
    \begin{split}
             {\rm SP}_2&=\frac{d_0}{6\cg^4}\sum_{P_1, P_2, P_3,P_4 \in \mg}\delta_{P_1, P_3}\delta_{P_4,P_1 P_2 P_1}=\frac{d_0}{6\cg^4}\sum_{P_1, P_2,P_4 \in \mg}\delta_{P_4,P_1 P_2 P_1}\\
             &\overset{(a)}{=}\frac{d_0}{6\cg^4}\sum_{P_1, P_2,P_4 \in \mg}\delta_{P_4,P_2}=\frac{d_0}{6\cg^3}\sum_{P_2,P_4 \in \mg}\delta_{P_4,P_2}=\frac{d_0}{6\cg^2}\,,
    \end{split}
\end{equation}
where in $(a)$ we exploited the fact that $\mg$ is an abelian subgroup of Pauli operators, and that Pauli operators square to the identity. The calculations for ${\rm SP}_3$ and ${\rm SP}_4$ yield the same result, whereas
\begin{equation}
    \begin{split}
        {\rm SP}_5&=\frac{1}{6\cg^4}\sum_{P_1, P_2, P_3,P_4 \in \mg}\delta_{P_2 P_4 P_3, P_1}=\frac{d_0}{6\cg^4}\sum_{P_2, P_3,P_4 \in \mg}1=\frac{1}{6\cg}
    \end{split}
\end{equation}
and the same result holds for ${\rm SP}_6$. Finally, we get
\begin{equation}
   \begin{split}
        {\rm SP}_0(\hi_G)&=\tr(Q^0 \Pi_{\sym}^0 \Pi_G ^{\ot 4})=\frac{1}{6}\left(\frac{d_0 ^2}{\cg^3}+3\frac{d_0}{\cg^2}+\frac{2}{\cg}\right)\\
        &=\frac{1}{6\cg}\left(\frac{d_0 ^2}{\cg^2}+3\frac{d_0}{\cg}+2\right)\,.
   \end{split}
\end{equation}
Conversely, the value of $\tr(Q^G \Pi_{\sym} ^G)$ reads
\begin{equation}
    \begin{split}
        \tr(Q^G \Pi_{\sym} ^G)&=\frac{1}{6}\left(d_G ^2+3 d_G +2\right)= \frac{1}{6}\left(\frac{d_0 ^2}{\cg^2}+3\frac{d_0}{\cg}+2\right)\,,
    \end{split}
\end{equation}
where we substituted $d_G=d_0/\cg$, which, in turn, implies $d_G/d_0=1/\cg$. Confronting, we then have 
\begin{equation}
    \tr(Q^0 \Pi_{\sym}^0 \Pi_G ^{\ot 4})=\frac{d_G}{d_0} \tr(Q^G \Pi_{\sym} ^G)
\end{equation}
which was the zero-gap condition shown in Eq. \eqref{0gap}.

\section{Stabilizer Entropy for qudits}\label{appHW}
To deal with a generic qudit systems, we shortly review the main ingredients of the generalization of Stabilizer Entropy for qudits.

Consider an $l$-dimensional Hilbert space $\mathcal{H}\simeq\mathbb{C}^l=\text{span}\{\ket{i},i\in Z_l\}$ and the space of linear operators $\mathcal{L}(\mathcal{H})$: we define the boost and shift operators $X,Z\in \mathcal{L}(\mathcal{H})$ as
\begin{align}
X\ket{j}&=\ket{j\oplus_l 1} \\
Z\ket{j}&=\omega^j\ket{j}\,,
\end{align}
with $\omega=e^{\frac{2\pi i}{l}}$.
By definition $X^l=Z^l=\Id$ and the commutation relation is 
\be 
XZ=\omega^{-1}ZX\implies X^pZ^q=-\omega^{-pq}Z^qX^p 
\ee 
We define a basis for $\mathcal{L}(\mathcal{H})$ as the set of the $l^2$ Heisenberg-Weyl operators
\be
D_{(p,q)}=\tau^{-pq}Z^{p}X^{q} \quad \text{with}\quad \tau=e^{\frac{i}{l}\pi} \quad p,q\in Z_d 
\ee
with product given by 
\be 
D_{(a,b)}D_{(c,d)}=\tau^{bc-ad}D_{(b+c,a+d)}
\ee 
and orthogonality relation
\be 
\tr (D_{(a,b)}D^\dag_{(c,d)})=l\delta_{a,c}\delta_{b,d}.
\ee
We define the discrete Heisenberg-Weyl group as the group generated by such operators:
\be
\tilde{\mathcal{D}}^{(l)}_1=\left<\{D_{(p,q)}\}\right>=\{\tau^aD_{(p,q)}\}\quad \text{with}\quad a,p,q\in Z_l
\ee

For a $n$-qudit system 
\be
\tilde{\mathcal{D}}^{(l)}_n=\{\tau^bD_{(p_1,q_1)}\otimes D_{(p_2,q_2)}\otimes \ldots D_{(p_n,q_n)}\}\quad \text{with}\quad b,p_i,q_i\in Z_l\quad \forall i\in\{1,\ldots,n\}
\ee

We can define the Clifford group for n qudit as the normalizer of the discrete HW group
\be 
C_n^{(l)}:=\{C\in \mathcal{U}(\hil)\,\,:\,\, C\tilde{\mathcal{D}}^{(l)}_nC^{\dag}=\tilde{\mathcal{D}}^{(l)}_n\} 
\ee 

The stabilizer entropies $M_{\alpha}$ can thus be generalized for a $n$-qudit system and, in particular, if we define the quotient 
\be 
\mathcal{D}^{(l)}_n=\frac{\tilde{\mathcal{D}}^{(l)}_n}{\tau^b \Id} \ee 
and the projector $Q=\frac{1}{d^2}\sum_{D_{(p,q)}\in \mathcal{D}^{(l)}_n}(D_{(p,q)}\otimes D_{(p,q)}^{\dag})^{\otimes 2}$
the usual definition of the Linear Stabilizer entropy still holds for qudits:
\be M_{lin(\psi)}=1-d\tr(Q\psi^{\ot 4}) \ee

\section{Calculation of the $Z_l$ SE gap}\label{appzl}
Recall that 
\be 
Q=\frac 1{d^2}\sum_{D\in\mathcal{D}}(D\ot D^{\dag})^{\ot 2} \quad \mlin(\psi)=1-d\tr(Q\psi^{\ot 4})\ee 
To calculate the magic gap we need to average over the states in the Hilbert space, hence the terms we need to calculate is
\be SP_0(H_G):=\tr(Q^0\Pi_{\sym}^0\Pi_G^{\ot 4}) \ee 
with $\Pi_G^{\ot 4}:=\frac{1}{\absval{G}^4}\sum_{D_1,D_2,D_3,D_4}D_1\ot D_2\ot D_3 \ot D_4$.
We first calculate the general term $\tr(Q\Pi_{\sym}\Pi_G^{\ot 4})$ and then obtain the expression for $tr(Q\Pi_{\sym})$ by setting the HW operators in the projector all equal to the identity. 
\be 
\tr(Q\Pi_{\sym}\Pi_G^{\ot 4})=\frac{1}{24 d^2}\sum_{\pi\in S_4}\sum_{D\in \mathcal{D}}\sum_{D_1,D_2,D_3,D_4\in \mathcal{G}}\tr\left[(D_1D)\ot(D_2D^{\dag})\ot(D_3D)\ot (D_4D^{\dag})T_{\pi}\right]
\ee 
In the following we calculate the $24$ traces for all the permutations in $S_4$ using the relation
\be \tr(\bigotimes_iA_iT_{\pi})=\tr(\prod_iA_{\pi(i)}) \ee 
We show some of the calculations as an example
\ba 
        &\bullet &({\rm id})\\
        &\sum_D&\tr(D_1D)\tr(D_2D^{\dag})\tr(D_3 D)\tr(D_4D^{\dag})= d^4\sum_D\delta_{D,D_1^{\dag}}\delta_{D,D_2}\delta_{D,D_3^{\dag}}\delta_{D,D_4}\\
        &=&d^4\delta(D_1^{\dag}=D_2=D_3^{\dag}=D_4) \\
        &\bullet &(12)\\
        &\sum_D&\tr(D_1 D D_2 D^{\dag}) \tr(D_3 D)\tr(D_4D^{\dag})= d^2\sum_D\delta_{D,D_3^{\dag}}\delta_{D,D_4} \tr(D_1 D D_2 D^{\dag})\\ &=&d^2\delta_{D_3^{\dag},D_4}\tr(D_1D_3^{\dag}D_2D_3)=d^3\delta_{D_3^{\dag},D_4}\delta_{D_1,D_3^{\dag},D_{2}^{\dag}D_3}\\
        &\bullet &(123) \\
        &\sum_D&\tr(D_3DD_1DD_2D^{\dag})\tr(D_4D^{\dag}) =d\sum_D\delta_{D_4,D}\tr(D_3DD_1DD_2D^{\dag}) \\ &=&d\tr(D_3D_4D_1D_4D_2D_4^{\dag})=d^2\delta_{D_3,D_4D_2^{\dag}D_4^{\dag}D_1^{\dag}D_4^{\dag}} \\
    \ea
    To deal with the $4$-cycles and the couples of swap operators we use the following relation:
    \be D^{\dag}D_iD=K(D,D_i)D_i \ee 
    where the phases $K(D,D_i)=\frac 1d\tr(D_i^{\dag}DD_iD^{\dag})$ compose according to the following expression:
    \be K(D,D_i)K(D,D_j)=K(D,D_iD_j) \ee 
    Using this expression, we get
    \ba \nonumber
        &\bullet &(1234)\\\nonumber
        &\sum_D&\tr(D_4D^{\dag}D_1DD_2D^{\dag}D_3D)= \sum_D K(D,D_1)K(D,D_3)\tr(D_4D_1D_2D_3)\\ 
        &=&d\sum_D K(D,D_1D_3)\delta_{D_4,D_3^{\dag}D_2^{\dag}D_1^{\dag}}\\\nonumber
        &\bullet &(12)(34)\\\nonumber
        &\sum_D&\tr(D_2D^{\dag}D_1D)\tr(D_4D^{\dag}D_3D)=d^2\sum_DK(D,D_1D_3)\delta_{D_2,D_1^{\dag}}\delta_{D_4,D_3^{\dag}} 
    \ea
    In the following, we list the values of all the traces:
    \ba
    \nonumber (\rm id)&=&d^4\delta(D_1^{\dag}=D_2=D_3^{\dag}=D_4) \\ \nonumber
    (12) &=& d^3 \delta_{D_3^{\dag},D_4}\delta_{D_1,D_3^{\dag}D_2^{\dag}D_1^{\dag}} \\ \nonumber
     (13) &=& d^3 \delta_{D_4,D_2}\delta_{D_1,D_2^{\dag}D_3^{\dag}D_2^{\dag}} \\ \nonumber
      (14) &=& d^3 \delta_{D_3^{\dag},D_2}\delta_{D_1,D_2D_4^{\dag}D_2^{\dag}} \\ \nonumber
       (23) &=& d^3 \delta_{D_1^{\dag},D_4}\delta_{D_2,D_1D_3^{\dag}D_1^{\dag}} \\ \nonumber
        (24) &=& d^3 \delta_{D_1^{\dag},D_3^{\dag}}\delta_{D_2,D_1^{\dag}D_4^{\dag}D_1^{\dag}} \\ \nonumber
         (34) &=& d^3 \delta_{D_1^{\dag},D_2}\delta_{D_3,D_1^{\dag}D_4^{\dag}D_1} \\ \nonumber
        (123) &=& d^2 \delta_{D_3,D_4D_2^{\dag}D_4^{\dag}D_1^{\dag}D_4^{\dag}} \\  
          (132) &=& d^2 \delta_{D_2,D_4^{\dag}D_1^{\dag}D_4^{\dag}D_3^{\dag}D_4} \\ \nonumber
           (142) &=& d^2 \delta_{D_2,D_1D_3^{\dag}D_1^{\dag}D_4^{\dag}D_1^{\dag}} \\ \nonumber
            (234) &=& d^2 \delta_{D_4,D_1D_3^{\dag}D_1^{\dag}D_2^{\dag}D_1^{\dag}} \\ \nonumber
             (243) &=& d^2 \delta_{D_4,D_1D_3^{\dag}D_1^{\dag}D_2^{\dag}D_1^{\dag}} \\ \nonumber
             (134) &=& d^2 \delta_{D_4,D_2^{\dag}D_3^{\dag}D_2^{\dag}D_1^{\dag}D_2} \\ \nonumber
              (124) &=& d^2 \delta_{D_4,D_3^{\dag}D_2^{\dag}D_3D_1^{\dag}D_3^{\dag}} \\ \nonumber
                (143) &=& d^2 \delta_{D_3,D_2^{\dag}D_1^{\dag}D_2D_4^{\dag}D_2^{\dag}} \\ \nonumber
                  (1234) &=& d \sum_D K(D,D_1D_3)\delta_{D_4,D_3^{\dag}D_2^{\dag}D_1^{\dag}} 
                  \ea 
    \ba \nonumber(1324) &=& d \sum_D K(D,D_3D_4D_3D_1)\delta_{D_4,D_2^{\dag}D_1^{\dag}D_3^{\dag}} \\ \nonumber
                   (1423) &=& d \sum_D K(D,D_1D_3D_2D_1)\delta_{D_3,D_1^{\dag}D_2^{\dag}D_4^{\dag}} \\ \nonumber
                    (1432) &=& d \sum_D K(D,D_3D_1)\delta_{D_2,D_1^{\dag}D_4^{\dag}D_3^{\dag}} \\ \nonumber
                     (1342) &=& d \sum_D K(D,D_1D_4D_1D_3)\delta_{D_2,D_3^{\dag}D_1^{\dag}D_4^{\dag}} \\ \nonumber
                      (1243) &=& d \sum_D K(D,D_3D_2D_3D_1)\delta_{D_2,D_4^{\dag}D_1^{\dag}D_3^{\dag}} \\ \nonumber
                       (12)(34) &=& d^2 \sum_D K(D,D_1D_3)\delta_{D_2,D_1^{\dag}}\delta_{D_4,D_3^{\dag}} \\ \nonumber
                         (13)(24) &=& d^2 \sum_D K(D,D_3D_2)\delta_{D^2,D_1^{\dag}D_3^{\dag}}\delta_{D^2,D_4D_2} \\ \nonumber
                           (14)(23) &=& d^2 \sum_D K(D,D_1D_3)\delta_{D_4,D_1^{\dag}}\delta_{D_2,D_3^{\dag}} 
    \ea 
We can now derive the general form for $\tr(Q\Pi_{\sym})$ for a $n$-qudit system by setting $D_i=\Id\,\, \forall i$. 
Looking at the 24 expressions above, all the $\delta$ in the identity, swaps and $3$-cycles return 1. Let us focus on the non-trivial case of $4$-cycles and double swaps. We will use the relation \cite{aharonov_2017interactiveproofsquantumcomputations} 
\be 
\sum_D K(D,D_i)=d^2\delta_{D_i,\Id} \ee 
to deal with the phases. 

We look at the following example: consider the contribution given by the cycle $(1234)$, i.e. 
\be 
d\delta_{D_4,D_3^{\dag}D_2^{\dag}D_1^{\dag}}\sum_D K(D,D_1D_3)=d\sum_D K(D,\id)=d^3 \ee
The same result holds for all the $4$-cycles. 

Let us look the double swaps:

    \ba  &(12)(34)&= d^2\delta_{D_2,D_1^{\dag}}\delta_{D_4,D_3^{\dag}}\sum_D K(D,D_1D_3)=d^2\sum_DK(D,\id)=d^4\nonumber\\ 
      &(14)(23)&=d^2\delta_{D_4,D_1^{\dag}}\delta_{D_2,D_3^{\dag}}\sum_D K(D,D_1D_3)=d^2\sum_DK(D,\id)=d^4 \\ 
    &(13)(24)&\sum_D\tr(D_3DD_1D)\tr(D_4D^{\dag}D_2D^{\dag})=\sum_D\tr(D^2)\tr({D^{\dag}}^2)=d^2\sum_D\delta_{D^2,\id} \nonumber
    \ea

For odd $l$, the only HW operator that squares to the identity is the identity itself. 

For even $l$ the counting is more complicated: for each slot of the n qudit string, we can either put $X^{\frac l2}$, $Z^{\frac l2}$, $X^{\frac l2}Z^{\frac l2}$ or the identity. Excluding the string $\id^{\ot n}$ we have $4^n-1$ HW operators such that $D^2=\id$. Hence, the contribution of the permutation $(13)(24)$ can be written as follows:
\be (13)(24)=d^2\left[1+(1-\text{mod}_2(l))+\text{mod}_2(l)\right]=d^2\left[4^n(1-\text{mod}_2(l))+\text{mod}_2(l)\right] \ee 

Collecting all the results, we can calculate $\tr(Q\Pi_{\sym})$ as
\be \tr(Q\Pi_{\sym})=\frac{1}{24}(3d^2+12d+8+4^n(1-\text{mod}_2(l))+\text{mod}_2(l)) \ee
For $l=2$, considering the relation $4^n=d^2,$ this formula reduces to
\be \tr(Q\Pi_{\sym})\frac 1{24}(3d^2+12d+8+d^2)=\frac{1}{24}(4d^2+12d+8)=\frac{(d+1)(d+2)}{6} \ee
which is the usual result of this quantity for a system of qubits \cite{Leone_2021_QCIQ,Leone_2022_SRE}.

Now, to calculate the second term in the definition of the magic gap, we need to perform the sum on all the $D_i$ in the group $\mathcal{G}$.
\begin{itemize}
    \item (Id) 

    \be \sum_{D_1,D_2,D_3,D_4}d_0^4\delta(D_1=D_2=D_3=D_4)=d_0^4\absval{G} \ee 
    \item (12)=(13)=(14)=(23)=(24)=(34)

    \be\sum_{D_1,D_2,D_3,D_4}d_0^3\delta_{D_3^{\dag},D_4}\delta_{D_1,D_3^{\dag}D_2^{\dag}D_3}=d_0^3\absval{G}^2 \ee 

    \item (123)=(132)=(142)=(124)=(134)=(143)=(234)=(243)

    \be d_0^2\sum_{D_1,D_2,D_3,D_4}\delta_{D_3,D_4D_2^{\dag}D_4^{\dag}D_1^{\dag}D_4^{\dag}}=d_0^2\absval{G}^3 \ee 

    \item (1234)=(1243)=(1324)=(1342)=(1423)=(1432)

    \be 
    d_0\sum_{D_1,D_2,D_3,D_4}\sum_D K(D,D_1D_3)\delta_{D_4,D_3^{\dag}D_2^{\dag}D_1^{\dag}}=d_0^3\sum_{D_1,D_2,D_3}\delta_{D_1D_3,\id}=d_0^3\absval{G}^2 \ee 

    \item (12)(34)=(14)(23) 

    \ba d_0^2\sum_{D_1,D_2,D_3,D_4}\sum_D K(D,D_1D_3)\delta_{D_2,D_1^{\dag}}\delta_{D_4,D_3^{\dag}}&=&d_0^2\sum{D_1,D_3}\sum_D K(D,D_1D_3) \\ &=&d_0^4\sum_{D_1,D_3}\delta_{D_1D_3,\id}=d_0^4\absval{G} \ea

    \item (13)(24) 

    \ba &d_0^2&\sum_D\sum_{D_1,D_2,D_3,D_4} K(D,D_3)K(D,D_4) \delta_{D^2,D_1^{\dag}D_3^{\dag}}\delta_{D^2,D_4D_2}= d_0^2\bigg(\sum_{D \in G}\sum_{D_1,D_2,D_3,D_4}\delta_{D^2,D_1^{\dag}D_3^{\dag}}\delta_{D^2,D_4D_2} \nonumber \\
    &+&\sum_{D\notin G}\sum_{D_1,D_2,D_3,D_4} K(D,D_3)K(D,D_4) \delta_{D^2,D_1^{\dag}D_3^{\dag}}\delta_{D^2,D_4D_2}\bigg)\nonumber \\
    &=& d_0^2\bigg(\sum_{D\in G}\sum_{D_2,D_3} + \sum_{D\notin G}\sum_{D_1,D_2,D_3,D_4} K(D,D_3)K(D,D_4) \delta_{D^2,D_1^{\dag}D_3^{\dag}}\delta_{D^2,D_4D_2}\bigg)
    \ea 
    The second term vanishes, as the first one returns $\sum_{D,D_2,D_3\in G}=\absval{G}^3$. So this last term yields the contribution 
    \be 
    (13)(24)=d_0^2\absval{G}^3 \ee
\end{itemize}

Collecting all the terms above, we have 
\be 
\tr(Q^0\Pi_{\sym}^0\Pi_G^{\ot 4})=\frac{1}{24}\left[\frac{3d_0^2}{\absval{G}^3}+\frac{12 d_0}{\absval{G}^2}+\frac{9}{\absval{G}}\right] \ee 
\be 
\tr(Q^G\Pi_{\sym}^G)=\frac{1}{24}\left[3d_G^2+(1-\text{mod}_2(l))(4^{n}-1)+12d_G+9\right] \ee
where $n$ is the number of sites in the lattice. 

Hence, the magic gap reads:
\ba 
\Delta M &=&d_G\exv_{U_G}\tr(Q_G\psi_{U_G}^{\ot 4}-d_0\exv_{U_G}\tr(Q^0\psi_{U_G}^{\ot 4})\nonumber \\
&=&d_G\binom{d_G+3}{4}^{-1}\tr(Q_G\Pi_{\sym}^G)-d_0\binom{d_G+3}{4}^{-1}\tr(Q^0\Pi_{\sym}^0\Pi_G^{\ot 4}) \nonumber \\
&=& \binom{d_G+3}{4}^{-1}\frac{1}{24}\bigg[d_G(3d_G^2+(1-\text{mod}_2(l))(4^{n}-1)+12d_G+9)-d_0\left(\frac{3d_0^2}{\absval{G}}+\frac{12d_0}{\absval{G}^2}+\frac{9}{\absval{G}}\right)\bigg] \nonumber \\ 
&=&\binom{d_G+3}{4}^{-1}\frac{1}{24}\left[3d_G^3+d_G(1-\text{mod}_2(l))(4^{n}-1)+12d_G^2+9d_G-3d_G^3-12d_G^2-9d_G\right] \nonumber \\
&=& \frac{(1-\text{mod}_2(l))}{(d_G+3)(d_G+2)(d_G+1)}(4^{n}-1) 
\ea

\end{document}